\documentclass[
    ,final            
  ]
  {aipproc}

\layoutstyle{8x11double}

\begin{document}

\title{Exclusive SUSY measurements and determination of SUSY
  parameters from LHC data}

\classification{04.65.+e, 12.60.Jv, 11.30.Pb}
\keywords      {Supersymmetry, LHC}

\author{Peter Wienemann for the ATLAS Collaboration}{
  address={Physikalisches Institut, Universit\"at Bonn, Nussallee 12,
    53115 Bonn, Germany} }

\begin{abstract}
A selection of exclusive SUSY measurements is presented which can be
performed provided SUSY will be discovered at the LHC. Such
measurements allow to determine the properties of supersymmetry and
thus help to pin down the underlying theoretical model. It is
described on an mSUGRA example point how sparticle masses can be
reconstructed from endpoints in mass spectra using early LHC
data. Finally it is shown how well mSUGRA model parameters can be
derived from these measurements.
\end{abstract}

\maketitle


\section{Introduction}

It is widely expected that the Large Hadron Collider (LHC) which
starts operation in autumn 2008 will uncover physics beyond the
present standard model of particle physics. Supersymmetry (SUSY) is
one of the most promising candidate for new physics. Among its virtues
are the potential to overcome the hierarchy problem, to provide a dark
matter candidate and make a unification of gauge coupling constants at
a high energy scale possible. If the SUSY mass scale is in the sub-TeV
range, already first LHC data will be sufficient to claim the
discovery of new physics. The subsequent challenge will be to
determine the properties of the observed new physics. Among the
questions seeking an answer are: Is it really SUSY? If yes, SUSY of
which type? How is SUSY broken? Finally everything should be destilled
to a set of Lagrangian parameters. To accomplish this, as many
observables as possible need to be studied. They include masses,
couplings and spins. In this report we will concentrate on techniques
to extract masses of SUSY particles using first LHC data, based on an
integrated luminosity of 1~fb$^{-1}$ in a favourable mSUGRA bulk
region scenario.

\section{SUSY mass measurements}

In the following we assume conserved $R$-parity. As a consequence
sparticles can only be produced pairwise and the lightest SUSY
particle (LSP) is stable. To evade cosmological problems, the LSP must
be a neutral, weakly interacting particle which escapes detection in
high-energy physics detectors. Due to two escaping LSPs in every SUSY
event, no mass peaks can be reconstructed and masses must be measured
by other means.

\begin{figure}
  \includegraphics[height=0.15\textheight]{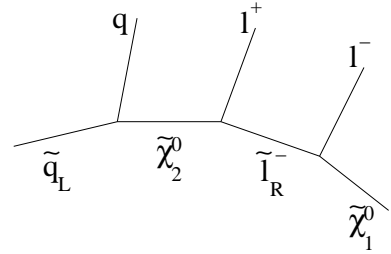}
  \caption{Prime example of a decay chain allowing a model independent
    reconstruction of the involved SUSY masses.}
  \label{fig:decaychain}
\end{figure}

The main source of mass information is provided by $\tilde{\chi}_2^0$
decays such as $\tilde{\chi}_2^0 \rightarrow \tilde{\ell}^{\pm}
\ell^{\mp} \rightarrow \tilde{\chi}_1^0 \ell^+ \ell^-$,
$\tilde{\chi}_2^0 \rightarrow \tilde{\chi}_1^0 Z^{(\ast)}$ and
$\tilde{\chi}_2^0 \rightarrow \tilde{\chi}_1^0 h \rightarrow
\tilde{\chi}_1^0 b \overline{b}$~\cite{ref:susymasses}. Provided that
a decay chain with at least three two-body decays can be identified,
the masses of the involved SUSY masses can be reconstructed in a
model-independent way.  As an example for such a case we assume that
the decay chain shown in Fig.~\ref{fig:decaychain} is open and
illustrate some important mass extraction techniques on this
example. This collection of possible measurements is based on studies
described in detail in~\cite{ref:csc}.

\subsection{Di-lepton mass}

First we consider the invariant mass spectrum of the two leptons
$m_{\ell \ell}$ from the decay chain in Fig.~\ref{fig:decaychain}. Due
to the scalar nature of the slepton, $m_{\ell \ell}$ exhibits a
triangular shape with a sharp drop-off at a maximal value $m_{\ell
  \ell}^{\mathrm{max}}$. The position of this endpoint depends on the
masses of the sparticles involved in the decay:
\begin{equation}
  m_{\ell \ell}^{\mathrm{max}} = m_{\tilde{\chi}_2^0} \sqrt{1 - \left(
    \frac{m_{\tilde{\ell}_R}}{m_{\tilde{\chi}_2^0}} \right)^2} \sqrt{1
    - \left( \frac{m_{\tilde{\chi}_1^0}}{m_{\tilde{\ell}_R}}
    \right)^2}.
\label{eq:endpoint}
\end{equation}
To suppress combinatorial SUSY and standard model (SM) background, the
endpoint is measured from the flavour subtracted di-lepton mass
distribution $N(\mathrm{e}^+\mathrm{e}^-)/\beta + \beta N(\mu^+\mu^-)
- N(\mathrm{e}^{\pm}\mu^{\mp})$. Here $N$ means the respective number
of events and $\beta$ is the ratio of the electron and muon
reconstruction efficiencies ($\beta \approx 0.86$).
\begin{figure}
  \includegraphics[height=0.2\textheight]{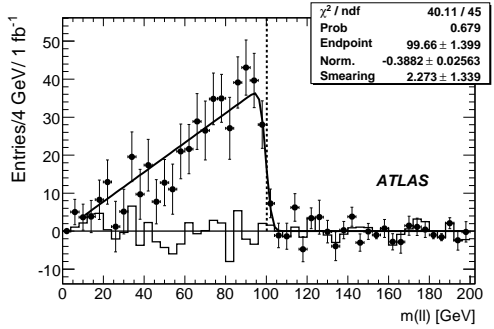}
  \caption{Flavour subtracted di-lepton mass spectrum for the ATLAS
    bulk region point SU3. The dashed line shows the expected endpoint
    position.}
  \label{fig:dilepton}
\end{figure}
Fig.~\ref{fig:dilepton} shows this mass distribution for the ATLAS
bulk region point SU3 ($\tan \beta = 6$, $M_0 = 100$~GeV, $M_{1/2} =
300$~GeV, $A_0 = -300$~GeV, sign$(\mu) = +1$). For an integrated
luminosity of 1~fb$^{-1}$, the position of the endpoint can be
reconstructed to be $(99.7 \pm 1.4 \pm 0.3)$~GeV where the first
uncertainty is statistical and the second one due to systematic
uncertainties from the lepton energy scale and $\beta$. This needs to
be compared to the nominal endpoint position for the studied benchmark
point which is located at $100.2$~GeV, indicated by the dashed line in
Fig.~\ref{fig:dilepton}. Apart from this simple triangular shape for a
single two-body decay also more complicated situations are possible
such as several superimposed triangles if more than one two-body decay
is open or even more general, SUSY-model dependent shapes if only
three-body decays are possible.

\subsection{Visible di-tau mass}

\begin{figure}
  \includegraphics[height=0.2\textheight]{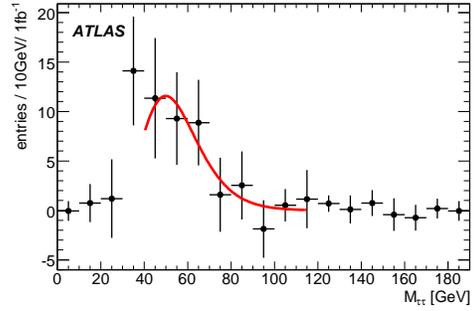}
  \caption{Charge subtracted visible di-tau mass spectrum for the
    ATLAS bulk region point SU3.}
  \label{fig:ditau}
\end{figure}

A similar analysis can also be performed if one replaces the electrons
and muons by taus. But due to the escaping neutrinos from the tau
decay, the visible di-tau mass distribution is not triangular anymore
(see Fig.~\ref{fig:ditau}). This complicates measuring the endpoint of
the spectrum. A rather robust strategy to evade this problem is to fit
a suitable function to the trailing edge of the visible di-tau mass
spectrum and use the inflection point as a endpoint sensitive
observable. The relation between endpoint and inflection point can be
established by a simple (rather model-independent) calibration
procedure. Fig.~\ref{fig:ditau} shows the charge subtracted visible
di-tau mass distribution $N(\tau^+\tau^-) - N(\tau^{\pm}\tau^{\pm})$
which is used to suppress background from fake taus and combinatorical
background. The reconstructed endpoint position for the SU3 benchmark
point is $(102 \pm 17^{\mathrm{stat}} \pm 5.5^{\mathrm{syst}} \pm
7^{\mathrm{pol}})$~GeV to be compared with the nominal value of 98
GeV. The last uncertainty is introduced by the SUSY-model dependent
polarisation of the two taus.

\subsection{Leptons $+$ jet mass endpoints}

By including the jet produced in association with the
$\tilde{\chi}^0_2$ in the $\tilde{q}_L$ decay (see
Fig.~\ref{fig:decaychain}), several other endpoints of measurable mass
combinations are possible: $m_{\ell q(\mathrm{low})}^{\mathrm{max}}$,
$m_{\ell q(\mathrm{high})}^{\mathrm{max}}$, $m_{\ell \ell
  q}^{\mathrm{min}}$ and $m_{\ell \ell q}^{\mathrm{max}}$. 
\begin{table}[h]
\begin{tabular}{lcc}
\hline
    \tablehead{1}{l}{t}{Endpoint}
  & \tablehead{1}{c}{t}{SU3 truth\\(GeV)}
  & \tablehead{1}{c}{t}{Measured\\(GeV)}\\
\hline
$m_{\ell q(\mathrm{low})}^{\mathrm{max}}$ & 325 & 333 $\pm$ 6 $\pm$ 6 $\pm$ 8\\
$m_{\ell q(\mathrm{high})}^{\mathrm{max}}$ & 418 & 445 $\pm$ 11 $\pm$ 11 $\pm$ 11\\
$m_{\ell \ell q}^{\mathrm{min}}$ & 249 & 265 $\pm$ 17 $\pm$ 15 $\pm$ 7\\
$m_{\ell \ell q}^{\mathrm{max}}$ & 501 & 517 $\pm$ 30 $\pm$ 10 $\pm$ 13\\
\hline
\end{tabular}
\caption{Reconstructed endpoint positions for leptons$+$jet masses for
  an integrated luminosity of 1~fb$^{-1}$ in the SU3 bulk point
  scenario. Given uncertainties are statistical, systematic and jet
  energy scale related, respectively.}
\label{tab:edges}
\end{table}
The label min/max denotes a lower/upper endpoint of the spectrum. The
index low/high indicates that the minimum/maximum of the two masses
$m_{\ell^+ q}$ and $m_{\ell^- q}$ is used. The reconstructed values
for these quantities are summarised in Tab.~\ref{tab:edges} for an
integrated luminosity of 1~fb$^{-1}$ in case of the bulk region point
SU3.  There are several reasons for the larger uncertainties on the
positions of these endpoints compared to the $m_{\ell
  \ell}^{\mathrm{max}}$ measurement. First, the endpoints are not as
pronounced as in the $m_{\ell \ell}$ case. Second, some distributions
suffer from remaining background close to the endpoint spoiling the
measurement and third, due to the involved jet, these measurements are
subject to the jet energy scale uncertainty which is an order of
magnitude larger than the lepton energy scale uncertainty.

\subsection{$\tilde{q}_R$ mass reconstruction}

Information about the $\tilde{q}_R$ mass can be gathered by exploiting
the decay $\tilde{q}_R \rightarrow \tilde{\chi}_1^0 q$. The endpoint
of the stransverse mass~\cite{ref:StransverseMass} distribution as
function of the $\tilde{\chi}_1^0$ mass hypothesis establishes a
relationship between the $\tilde{q}_R$ and the $\tilde{\chi}_1^0$
masses. In the SU3 scenario, the stransverse mass endpoint is
reconstructed to be 590~$\pm$~9(stat)~$^{+13}_{-6}$~GeV for the
nominal $\tilde{\chi}_1^0$ mass assuming 1~fb$^{-1}$ integrated
luminosity which needs to be compared with the nominal value of
637~GeV.

\subsection{Mass determination from edge positions}

The positions of the various endpoints are a function of the masses of
the sparticles involved in the respective decays. Since the inversion
of the endpoint formulae is only possible for a few specific cases,
the masses are extracted numerically by a $\chi^2$ fit. The
reconstructed masses using the di-lepton and the leptons$+$jet
endpoints in the SU3 scenario for an integrated luminosity of
1~fb$^{-1}$ is summarised in Tab.~\ref{tab:masses}.
\begin{table}[h]
\begin{tabular}{lcc}
\hline
    \tablehead{1}{l}{t}{Mass}
  & \tablehead{1}{c}{t}{SU3 truth\\(GeV)}
  & \tablehead{1}{c}{t}{Reconstructed\\(GeV)}\\
\hline
$m_{\tilde{\chi}_1^0}$ & 118 & 88 $\pm$ 60 $\mp$ 2\\
$m_{\tilde{\chi}_2^0}$ & 219 & 189 $\pm$ 60 $\mp$ 2\\
$m_{\tilde{\ell}_R}$ & 155 & 122 $\pm$ 61 $\mp$ 2\\
$m_{\tilde{q}_L}$ & 634 & 614 $\pm$ 91 $\pm$ 11\\ \hline
$m_{\tilde{\chi}_2^0} - m_{\tilde{\chi}_1^0}$ & 100.7 & 100.6 $\pm$ 1.9 $\mp$ 0 \\
$m_{\tilde{\ell}_R} - m_{\tilde{\chi}_1^0}$ & 37.6 & 34.2 $\pm$ 3.8 $\mp$ 0.1\\
$m_{\tilde{q}_L} - m_{\tilde{\chi}_1^0}$ & 516.0 & 526 $\pm$ 34 $\pm$ 13\\
\hline
\end{tabular}
\caption{Reconstructed sparticle masses for an integrated luminosity
  of 1~fb$^{-1}$ in the SU3 bulk point scenario.Given uncertainties
  are statistical and jet energy scale related. $\mp$ indicates an
  anti-correlation with the jet energy scale variation.}
\label{tab:masses}
\end{table}
The reconstructed masses are highly correlated with
$m_{\tilde{\chi}_1^0}$ which is only poorly contrained by the endpoint
measurements. Therefore the precision on the absolute mass values is
rather moderate. Significantly higher accuracy is obtained for the
mass differences between the various masses and the $\tilde{\chi}_1^0$
mass.

\section{mSUGRA parameter fits}

Apart from the model-independent mass reconstruction one also would
like to test the consistency of the data with various models and to
determine their parameters. To accomplish this, several programs are
available exploiting different parameter estimation techniques (see
e.~g.~\cite{ref:Fittino}, \cite{ref:SFitter}).
\begin{table}[h]
\begin{tabular}{lcc}
\hline
    \tablehead{1}{l}{t}{Parameter}
  & \tablehead{1}{c}{t}{SU3 truth}
  & \tablehead{1}{c}{t}{Fitted}\\
\hline
$\tan\beta$ & 6 & 7.4 $\pm$ 4.6\\
$A_0$ (GeV) & $-$300 & 445 $\pm$ 408\\
$M_0$ (GeV) & 100 GeV & 98.5 $\pm$ 9.3\\
$M_{1/2}$ (GeV) & 300 GeV & 317.7 $\pm$ 6.9\\
\hline
\end{tabular}
\caption{Fitted mSUGRA parameters for an integrated luminosity of
  1~fb$^{-1}$ in the SU3 bulk point scenario.}
\label{tab:parameters}
\end{table}
Tab.~\ref{tab:parameters} contains a summary of the precision with
which the mSUGRA parameters can be determined using the endpoint
measurements mentioned in the previous sections. Even with 1~fb$^{-1}$
the scalar and the gaugino mass parameter can already be determined to
better than 10~\%. For $\tan \beta$ and the trilinear coupling
parameter only an $\mathcal{O}(100~\%)$ measurement is feasible. But
as soon as Higgs sector information will become available with higher
luminosity the uncertainty on latter parameters will drop sizably.

\section{Conclusions}

Provided sparticles masses are in the sub-TeV regime, already first
LHC data will allow to perform a rough SUSY spectroscopy. First checks
of high-scale unification models are also feasible. Higher precision
and more difficult measurements follow as the integrated luminosity
increases.



\end{document}